\documentstyle[preprint,aps]{revtex} 
\tightenlines
\begin{document}


\title{Spin-orbit splitting in non-relativistic and relativistic 
self-consistent models}

\author{M. L\'opez-Quelle,$^{1,}$\thanks{e-mail address: lopezm@besaya.unican.es}
  {N. Van Giai},$^2$ {S. Marcos},$^3$ and {L.N. Savushkin}$^4$}
\address{$^1$Universidad de Cantabria, Departamento de F{\'\i}sica Aplicada, 39005 Santander, 
Spain}
\address{$^2$ Institut de Physique
Nucl\'eaire, F-91406 Orsay Cedex, France}
\address{$^3$Universidad de Cantabria, 
Departamento de F{\'\i}sica Moderna, 39005 Santander, 
Spain} 
\address{$^4$St.-Petersburg Institute for Telecommunications,
Department of Physics, 191065 St. Petersburg, Russia}

\maketitle

\begin{abstract}
The splitting of single-particle energies between spin-orbit partners in 
nuclei is examined in the framework of different self-consistent approaches, 
non-relativistic as well as relativistic. Analytical expressions of 
spin-orbit potentials are given for various cases. Proton spin-orbit 
splittings are calculated along some isotopic chains (O, Ca, Sn) and 
they are compared with existing data. It is found that the isotopic 
dependence of the relativistic mean field predictions is similar to 
that of some Skyrme forces while the relativistic Hartree-Fock approach 
leads to a very different dependence due to the strong non-locality.   


\end{abstract}


\newpage
\section{Introduction}
One often asked question about the relativistic approach to nuclear
structure is: what are the main physical effects that a relativistic
approach can handle better than a non-relativistic approach? Among various
arguments one feature seems to emerge quite obviously, namely the fact that,
in a description based on the Dirac equation the nucleon spin degree of
freedom is incorporated very naturally. Therefore, nuclear properties
related to the spin and particularly spin-orbit effects in nuclei should
come out from a relativistic description without any special adjustments
whereas a non-relativistic approach would need ad hoc parameters. This was
indeed so in the Dirac phenomenology analysis of spin observables measured
in medium energy nucleon-nucleus scattering\cite{Clark}. The interplay of
two deep phenomenological potentials, the attractive scalar field and the 
repulsive vector field gives the right magnitude for the central and
spin-orbit optical potentials. 

However, if one tries to calculate microscopically the mean field
experienced by a nucleon bound to a nucleus and predict spin-orbit
splittings, it is not so clear how successful will be the various
relativistic models. In this work, we would like to analyze the spin-orbit
splittings calculated in different self-consistent models: non-relativistic
Hartree-Fock with Skyrme-type forces, relativistic mean field theory (RMFT) and
relativistic Hartree-Fock (RHF). Of course, a mean field approach like 
Hartree-Fock is designed for reproducing global ground state properties 
like total binding energies or densities but it is not supposed to 
describe correctly single-particle spectra because core polarization 
effects are known to modify importantly single-particle 
energies\cite{HS76,BG80}. However, the core polarization effects should 
affect essentially in the same way the two members of a spin-orbit 
doublet if they are both below, or both above 
the Fermi level\cite{RU98}. Therefore, it is 
legitimate to study the question of spin-orbit splittings in 
Hartree-Fock or mean field frameworks for nuclei where both spin-orbit 
partners are occupied.

We carry out this study by looking at the evolution of the spin-orbit 
splittings of a $nlj_>-nlj_<$ proton pair of states along an isotopic 
chain. The predictions of different models are compared with experimental 
values when they are available. The difficulties of such comparisons lie 
in the fact that the experimental information is scarce and it suffers
sometimes of large uncertainties. This points to the necessity of having 
more data on spin-orbit splittings if one aims at more quantitative 
conclusions about the validity of different models concerning their 
spin-orbit properties. Within these limitations, our analysis will be 
done for the Ca, O and Sn isotopic chains.

The outline of the paper is as follows. In Section II and in the 
Appendix we derive the
spin-orbit potentials corresponding to the different models:
Skyrme-Hartree-Fock (SHF), RMFT and RHF. In Section III we discuss the 
results for the $^{40}$Ca-$^{48}$Ca case 
as well as those concerning the O and Sn isotopes.
Concluding remarks are made in Section IV.

\section{Formalism}

Let us derive the main expressions for the spin-orbit potentials in 
non-relativistic and relativistic approaches. In either case the general 
method is to rewrite the Hartree or Hartree-Fock equations for the 
single-particle states in a Schr{\"o}dinger-like form and to identify a 
central potential and a spin-orbit potential. These potentials are local 
in the cases considered here, but they are generally state-dependent. 

\subsection{Non-relativistic approach}

The SHF model is very simple because the Hartree-Fock equations for the 
single-particle wave functions $\phi_i$ and energies $\varepsilon_i$ take 
the form of a Schr{\"o}dinger equation with an effective mass $m^*({\bf r})$
 and a 
local, state-independent 
potential. All non-locality effects are described by this effective mass and consequently, the SHF equations contain also first 
derivatives of the wave function. It is natural to introduce an 
asymptotically equivalent wave function\cite{Be75}: 
\begin{eqnarray}
\tilde\phi_i ({\bf r}) & = & \Big(\frac {m^*({\bf r})}{m}\Big)^{-\frac 
{1}{2}} \phi_i ({\bf r})~,
\label{eq1}
\end{eqnarray}
which now satisfies a Schr{\"o}dinger equation with a constant mass $m$ 
and a state-dependent local potential:
\begin{eqnarray}
V(\varepsilon_i, {\bf r}) & = & V_0(\varepsilon_i, {\bf r}) + 
V_{LS}({\bf r}) {\bf l.s}~. 
\label{eq2}
\end{eqnarray}
The spin-orbit potential is (for spherical symmetry):
\begin{eqnarray}
V_{LS}(q, r) & = & \frac {m^{*}_q(r)}{m} \Big[ \frac {w}{r} 
[\rho'(r) + \rho_{q}{'}(r)] - \frac 
{1}{4r} [(t_1 x_1 + t_2 x_2) J(r) + (t_2 - t_1) J_{q} (r)] \Big]~,     
\label{eq3}
\end{eqnarray}
where $w$, $t_i$, $x_i$ are Skyrme force parameters, $\rho_q$ and $J_q$ 
are nucleon densities and spin densities ($q$= $n$ or $p$) with $\rho = 
\rho_n + \rho_p$ and $J = J_n + J_p$\cite{Be75}. The spin densities are 
practically zero in spin-saturated nuclei but they can give some 
contributions in spin-unsaturated ones. The above expression shows that 
$V_{LS}$ is surface peaked, and that proton spin-orbit splittings 
$\Delta_{LS}$ should 
have little isotopic dependence in the SHF model as we shall see in 
the next sections. One should, however, keep in mind that another 
contribution to $\Delta_{LS}$ comes from the energy-dependent central 
potential. One can easily check that the energy dependence of 
$V_0(\varepsilon_i, {\bf r})$ is entirely contained in a term $\big (
1-m^{*}({\bf r})/m \big) \varepsilon_i$\cite{Be75}. 

Recently, some attempts have been made in order to introduce more 
freedom in the spin-orbit term of the Skyrme parametrization with the 
aim of improving isotope shift predictions in the Pb region\cite{Rein95}. 
The authors of Ref.\cite{Rein95} modify directly the Skyrme energy 
functional so that the first term of Eq.(\ref{eq3}) becomes 
$w \rho{'}(r)/ r + w{'} \rho_{q}{'}(r) / r$ and the spin density terms are
dropped. It is found that a better 
description of data is reached when $w{'}$ is nearly 
$-w$ (parameter set SkI4), 
i.e., when the proton spin-orbit potential depends almost entirely on 
the derivative of the neutron density. 

\subsection{Relativistic approach}

Let us now calculate the spin-orbit potential in the general RHF case. 
The corresponding expressions for the RMFT case will be easily obtained by 
dropping all contributions corresponding to exchange (Fock) terms. 
Starting from the RHF equation for the 4-component spinors we shall 
obtain a Schr{\"o}dinger-like equation for the upper component which 
allows us to identify the central and spin-orbit potentials. We use the 
notations introduced in Ref.\cite{Bouy87}. 

We start from Eq.(C3) of 
Ref.\cite{Bouy87} for the spinor $(G_i, F_i)$:
\begin{eqnarray}
G{'}_i & = & (-\frac {\kappa_i}{r} - \Sigma_{T,i} - P_i) G_i + (m + E_i 
+\Sigma_{S,i} - \Sigma_{0,i} - Q_i ) F_i \nonumber \\ 
F{'}_i & = &  (m - E_i 
+\Sigma_{S,i} + \Sigma_{0,i} + R_i ) G_i 
+ (\frac {\kappa_i}{r} + \Sigma_{T,i} + S_i) F_i~,
\label{eq4}
\end{eqnarray}
where the $\Sigma$'s are the direct (Hartree) self-energies whereas 
$P,Q,R,S$ are related to the exchange (Fock) contributions. We shall 
drop the state indices $i$ from now on. The first step is to eliminate 
the lower component $F$ and obtain a second order differential equation 
for the upper component $G$:
\begin{eqnarray}
\Lambda G{''} + \alpha_1 G{'} + \alpha_0 G & = & 0~, 
\label{eq5}
\end{eqnarray}
where
\begin{eqnarray}
\Lambda & = & (m + E + \Sigma_S - \Sigma_0 - Q)^{-1}~,\nonumber \\
\alpha_1 & = & \Lambda{'} + \Lambda (P - S)~,\nonumber \\
\alpha_0 & = & \Lambda{'} (\frac {\kappa}{r} + \Sigma_T + P) + \Lambda 
(-\frac {\kappa}{r^2} + \Sigma{'}_T + P{'}) \nonumber \\
& & - (M-E+\Sigma_S + \Sigma_0 
+R) - \Lambda (\frac {\kappa}{r} + \Sigma_T + P)(\frac {\kappa}{r} + 
\Sigma_T + S)~.
\label{eq6}
\end{eqnarray}
Next, we look for a modified function $\tilde G$:
\begin{eqnarray}
G & = & \lambda \tilde G~,
\label{eq7}
\end{eqnarray}
such that $G$ and $\tilde G$ are identical asymptotically while the 
differential equation satisfied by $\tilde G$ contains no first 
derivative. This condition determines $\lambda$ to be:
\begin{eqnarray}
\lambda & = & C \Lambda^{-\frac {1}{2}} e^{-\frac {1}{2}\int (P-S) dr}~.
\label{eq8}
\end{eqnarray}
where the constant  C is determined from the condition $\int {\tilde G ^2}
dr=1$.
 In the limit of RMFT the exchange quantities $P$ and $S$ are zero and 
one recovers the familiar result $\lambda = \Lambda^{-\frac {1}{2}}$. 

We can now write the differential equation satisfied by $\tilde G$ in 
the form:
\begin{eqnarray}
-\frac {\hbar^2}{2m}\Big( \tilde G{''} + \big[\frac 
{\lambda{''}}{\lambda} - \frac {1}{2} (\frac {\alpha_1}{\Lambda})^2 + 
\frac {\alpha_0}{\Lambda}\big] \tilde G \Big) = 0~.
\label{eq9}
\end{eqnarray}
The potential in state $i$ can be identified from the coefficient of $\tilde 
G$. It is possible to separate out the spin-orbit part $V_{LS}$ of this 
potential by observing that the quantities $\Lambda, P, Q, R, S$ depend 
on $\kappa \equiv (2j + 1)(l-j)$ (see, e.g., Ref.\cite{Bouy87}) which in 
turn can be expressed in terms of $\langle {\bf l.s}\rangle$. The 
complete expressions for $V_{LS}$ are given in the Appendix.

\section{EVOLUTION OF SPLITTINGS ALONG ISOTOPIC CHAINS}

We now turn to an analysis of spin-orbit splittings in some finite 
nuclei. We concentrate on the quantities:
\begin{eqnarray}
\Delta(nlj_{<}-nlj_{>}) \equiv \varepsilon_{nlj_{<}} - 
\varepsilon_{nlj_{>}}~.
\label{eq10}
\end{eqnarray}

\subsection{Ca isotopes}

In Ca isotopes, the experimental situation concerning the $1d_{5/2}$
and $1d_{3/2}$ proton states is not very accurately established.
In $^{40}$Ca, the generally accepted value of $1d_{3/2}$ single-particle 
energy is $\varepsilon_{1d3/2}=-8.3$ MeV 
\cite{Mat65}, \cite{Ray79}, \cite{Oros96}
 However, the values of $\varepsilon_{1d5/2}$ are 
much more dispersed: -15.1 MeV \cite{Swif66}, -15.5 MeV \cite{Tyren66},
-16 MeV \cite{Ray79}, -14.3 MeV \cite{Oros96}. Note that the most recent 
values of Ref. \cite{Oros96} have been obtained as the centroid energies 
of the single-particle spectroscopic strength distributions:

\begin{eqnarray}
\varepsilon_{nlj} \equiv \frac {\sum_i E_i S_i}{\sum_i S_i} 
\label{eq11}
\end{eqnarray}
and therefore, they take into account the strong fragmentation of the 
single-particle states. The fragment energies $E_i$ and 
spectroscopic factors $S_i$ were taken from Refs. 
\cite{NNDC},\cite{End90},\cite{You70},\cite{Dev81},\cite{Eck90} and \cite{Dol76}.
The corresponding experimental spin-orbit splittings of $^{40}$Ca are
plotted on Fig. 1.
In $^{48}$Ca, the experimental proton energies $\varepsilon_{1d3/2}$
are respectively -16.2 MeV \cite{Ray79} and -15.5 MeV \cite{Oros96}
whereas the $\varepsilon_{1d5/2}$ energies are -21.5 MeV \cite{Ray79}
and -20.5 MeV \cite{Oros96}. This gives the spin-orbit energies in
$^{48}$Ca shown in Fig. 1.

At this point we must mention that some theoretical papers, e.g.,
Ref. \cite{Bouy87} quote a somewhat smaller value for the experimental
$\Delta_{S.O.}(1d3/2-1d5/2)$ in $^{48}$Ca based on an earlier paper
\cite{Camp72} but the experimental source remains uncertain.

 In Fig. \ref{fig1} we display the results calculated with 
the following models: SHF with a standard SIII force and with the SkI4 
parametrization of Ref.\cite{Rein95}, RMFT with the 
commonly used NL-SH \cite{NLSH} and NL3 \cite{NL3} parametrizations, 
RHF with ($\sigma$, $\omega$, $\rho$, $\pi$) mesons (model e) of 
Ref.\cite{Bouy87}). 
The two parametrizations NL-SH and NL3 have been chosen as the most 
representative and successful in describing nuclear 
ground states in a wide range of nuclei.

All calculated results for $\Delta(1d_{3/2}-1d_{5/2})$ show
 a linear A-dependence between $A$=40 to $A$=48. 
However, whereas the  values of SHF and RMFT models 
decrease only slightly  those of the  RHF model exhibit a 
large decrease.                                              

The relatively small variations of the SIII results can be understood by
examining the $V_{LS}$ potential of Eq.(\ref{eq3}).
When going from $^{40}$Ca to $^{48}$Ca one acquires extra neutron
densities $\rho_{7/2}(r)$ and $J_{7/2}(r)$
due to the filling of the $1f_{7/2}$ orbital while the core neutron and proton
densities do not change much. For the SIII parametrization, $x_1$ =
$x_2$ = 0 so that $J_{7/2}(r)$ gives no contribution to $V_{LS}(p,r)$
whereas $\rho{'}_{7/2}(r)$ yields a contribution with a node at the
surface, i.e., a small effect on $\Delta(1d_{3/2}-1d_{5/2})$. The same analysis
holds for the SkI4
results. For the RMFT case the interpretation of results is also simple. 
The spin-orbit potential reduces to the first term of Eq.(\ref{eqa1}), 
i.e., $-2 \Lambda{'}/\Lambda r$. With the definition of $\Lambda$ given 
by Eq.(\ref{eq6}) it is seen that the variation of spin-orbit potential 
is obtained by folding the derivative of the scalar and vector $1f_{7/2}$ 
densities with the (short-ranged) $\sigma$ and $\omega$ form factors
and therefore, it has a node near the surface and the variation 
of spin-orbit splitting must be small. 
We should note, however, that the central potential  that one can identify from
Eq.(\ref{eq9}) has some state dependence due to the effective mass $m^*=m+\Sigma_S$ and
this effect can also contribute to the energy splitting. Furthermore, the
kinetic energy and the central potential contribute also to this splitting
through the j-dependence of the radial part in the wave function.

The RHF results are more 
difficult to interpret. 
 From Eq.(\ref{eq9}) we identify the kinetic energy $T$, central 
potential $V_0$ and spin-orbit potential $V_{LS}$ such that:
\begin{eqnarray}
\langle T \rangle_i + \langle V_0 \rangle_i + \langle V_{LS} \rangle_i  & = & 
\varepsilon_i (1 + \frac {\varepsilon_i}{2 m})~.  
\label{eq12}
\end{eqnarray}
The expressions for $V_{LS}$ given in the Appendix are complicated. 
The non-locality of the original RHF mean field induces a 
strong state dependence in the local equivalent potentials and 
therefore, variations with $j_{>}$ and $j_{<}$ will be 
found in all three terms of 
Eq.(\ref{eq12}). This is illustrated in Table \ref{table2} where the 
values of the differences:
\begin{eqnarray}
\Delta T & \equiv & \langle T \rangle_{1d3/2} - \langle T 
\rangle_{1d5/2}\;, \nonumber \\ 
\Delta V_0 & \equiv & \langle V_0 \rangle_{1d3/2} - \langle V_0 
\rangle_{1d5/2}\;, \nonumber \\ 
\Delta V_{LS} & \equiv & \langle V_{LS} \rangle_{1d3/2} - \langle V_{LS} 
\rangle_{1d5/2}\;,
\label{eq13}
\end{eqnarray}
are shown. In $^{40}$Ca the 
calculated spin-orbit splitting is practically equal to 
$\Delta V_{LS}$, with a strong cancellation of the two other terms. In 
$^{48}$Ca $\Delta V_{LS}$ is only one third of $\varepsilon_{1d3/2}-
\varepsilon_{1d5/2}$~, the rest coming from an increase of 
$\Delta V_0$ and $\Delta T$. 

It is possible to analyze further the respective role of the isovector 
$\pi$ and $\rho$ mesons in the decrease of $\Delta(1d_{3/2}-1d_{5/2})$ 
from $^{40}$Ca to $^{48}$Ca. The key effect comes from the $\pi$ meson 
and the strong non-locality related to its light mass. If one 
artificially increases $m_{\pi}$ up to 600 MeV (and renormalize 
accordingly the $f_{\pi}$ coupling constant to keep reasonable 
single-particle levels) one tends to a more local situation and the 
strong decrease of $\Delta(1d_{3/2}-1d_{5/2})$ vanishes. The role of the 
$\rho$ meson can be studied by repeating the calculations without its 
contributions to the Fock terms. It is found that the decrease of 
$\Delta(1d_{3/2}-1d_{5/2})$ still remains.

\newpage
\subsection{Oxygen isotopes}

The O chain is particularly interesting because of the prospects of 
reaching experimentally new, unstable isotopes with a large neutron excess. 
We examine here the evolution of the $1p_{1/2}-1p_{3/2}$ splitting of 
proton levels. In 
Fig. \ref{fig2} we summarize the results obtained with SHF 
(SIII and SkI4 parametrizations), 
RMFT (NL-SH and NL3 parametrizations) and RHF (model $e)$ of Ref.\cite{Bouy87}). 
We note that the results of NL-SH and NL3 are rather similar for 
spin-orbit splittings in O isotopes. This behaviour remains also in 
the Sn isotopes (see next subsection). These two parametrizations
differ mostly by their predicted compression modulus K and therefore, 
this indicates that K has little effect on 
spin-orbit properties in the RMFT. 
One can distinguish three intervals ending at $A$=22, 24 and 28 which correspond 
to the filling of $1d_{5/2}$, $2s_{1/2}$ and $1d_{3/2}$ subshells, 
respectively. The trend in the $1d_{5/2}$ subshell resembles that of the 
Ca isotopes, namely a relatively small variation of 
$\Delta(1p_{1/2}-1p_{3/2})$ for the SHF and RMFT models and a large 
decrease for the RHF model. 
 A similar discussion as for the Ca isotopes 
can be done here. It would be interesting to measure experimentally 
the variations of spin-orbit splittings in the chain of Oxygen isotopes. 
 For the isotopes heavier than 
$A$=22 RMFT shows a maximum at $A$=24 followed by a decrease whereas 
for RHF the behaviour is opposite, with an increase of the spin-orbit 
splitting after $A$=22. 

\subsection{Sn isotopes}

In the chain of Sn isotopes proton spin-orbit splittings are 
experimentally known in several nuclei\cite{Sn-exp}. In Fig. \ref{fig3} 
are shown the calculated and measured values of $\Delta(2p_{1/2}-2p_{3/2})$ 
for protons, and in Fig. \ref{fig4} are displayed the calculated results for 
$\Delta(1f_{5/2}-1f_{7/2})$. From Fig. \ref{fig3} it can be seen that none of 
the models is able to reproduce the very small empirical values of 
$\Delta(2p_{1/2}-2p_{3/2})$. In the range A=112-124 the SIII results show 
relatively small variations whereas the 3 other models give large 
fluctuations. Of course, it would be more satisfactory to calculate these 
nuclei using a Hartree-Fock-BCS description. It is not clear, however, that 
the proton spin-orbit splittings would be significantly affected so as 
to bring them into agreement with the data. Indeed, in Ref.\cite{Shen96} a 
HF-BCS was performed for Sn isotopes using the effective interaction SkI4 and 
a density-dependent pairing force. The values of $\Delta(2p_{1/2}-2p_{3/2})$ 
thus obtained vary from 1.6 MeV in $^{112}$Sn to 2.0 MeV in $^{124}$Sn. As 
for the $1f_{5/2}-1f_{7/2}$ splitting, Fig. \ref{fig4} shows a wide range of 
predictions. For instance, in A=100 RHF is as low as 2.2 MeV while the other
models are in the 5-6 MeV range . It would be very interesting to have data on this 
$1f_{5/2}-1f_{7/2}$ case in order to shed light on this issue.

\section{CONCLUSION}

We have examined in this work some predictions of spin-orbit splittings in 
the framework of non-relativistic and relativistic mean field approaches. 
We selected, as a representative case of the non-relativistic self-consistent 
approach, the Skyrme-Hartree-Fock model because it is widely used and its 
analytic simplicity lends itself to an easy interpretation of the calculated 
splittings. Numerical results were obtained with two versions of the effective 
Skyrme interaction, the standard parametrization SIII and the more recently 
proposed SkI4 which contains more degrees of freedom in its spin-orbit part. 
On the relativistic side, the RMFT is also a successful model for describing 
nuclear ground states and we have examined its spin-orbit predictions with 
the often used non-linear versions NL-SH and NL3. Here also the analytic form of the 
one-body spin-orbit potential is simple enough to allow some insight into the 
numerical results. In addition, we have included in our study the RHF model 
of Ref.\cite{Bouy87} in order to examine the role of the pion in the 
spin-orbit splittings.

One conclusion which can be drawn is that there is no clear advantage of the 
RMFT over the non-relativistic Skyrme-Hartree-Fock approach. 
It is true that in RMFT, the spin-orbit properties come out together 
with the central part of the mean field whereas in the non-relativistic 
approach one has the freedom of one or more additional spin-orbit 
parameters. Nevertheless, it cannot be concluded that the RMFT 
spin-orbit splittings describe the data particularly well.
In Refs. \cite{BM83}, \cite{Rein89} an interesting formal connection is made between 
RMFT and Skyrme-HF, but we can see here that, at a quantitative level,
not only their spin-orbit predictions do differ, but even between SIII 
and SkI4 there are sizable differences in $\Delta_{S.O.}$

All models fail to reproduce the very small $2p_{1/2}-2p_{3/2}$ proton splittings 
in Sn isotopes and it does not seem that pairing correlations can improve 
this prediction. As for the RHF approach, the light pion mass produces 
strongly non-local Fock fields, i.e., a strong state dependence of the mean 
fields. Consequently, the splitting between spin-orbit partners is due not 
only to $\langle V_{LS}{\bf l.s} \rangle_{lj}$ but also to a large extent 
to the state dependence of the central potential. Thus, the evolution of the 
predicted splittings along an isotopic chain does not resemble that of less 
non-local models like RMFT or Skyrme. It appears that the isotopic dependence 
of $\Delta$ disagrees with the data in the case of RHF.

The conclusion made earlier \cite{Bouy87} that RHF describes well the
experimental $\Delta_{S.O.}$ both in $^{40}$Ca and $^{48}$Ca 
was too premature in view of the uncertainties in the experimental 
values.

Finally, the question of spin-orbit predictions in the framework of 
self-consistent theories is still open. Existing parametrizations of effective 
interactions, relativistic as well as non-relativistic, need further 
improvements. 
The RMF predictions seem to follow a common trend whereas the Skyrme-HF 
results may differ somewhat depending on the parametrizations, as it 
can be seen with SIII and SkI4 in Sn isotopes. Improving the parametrizations 
necessitates better comparisons with experiment.
In this respect, it would be very helpful to have more data 
on the evolution of spin-orbit splittings along isotopic chains.
Without better data, it is not possible at the moment to state that the 
spin-orbit problem is understood either by RMFT or by the non-relastivistic 
approach.

\acknowledgements

We would like to thank M. Bender and R. Wyss for their comments and A.M. Oros for the data on 
Ca isotopes. 

This work was supported in part by contract DGICYT PB97-0360 (Spain).

\begin{appendix}
\section{}

In this Appendix we give the expressions for $V_{LS}$ in the general case of 
RHF.
The special case of RMFT is easily obtained by setting all functions 
$P, Q, R, S$ to zero.
The notations are those of Ref.\cite{Bouy87}~.

Eq.(\ref{eq9}) is a Schr\"odinger-type equation for a state of angular 
momentum ($l,j$). We split the coefficient of $\tilde G$ into a centrifugal 
term, a central potential and a spin-orbit potential:
\begin{eqnarray}
\frac {\lambda{''}}{\lambda} - \frac {1}{2} (\frac {\alpha_1}{\Lambda})^2 + 
\frac {\alpha_0}{\Lambda} & \equiv & \frac {-l(l+1)}{r^2} - 
\frac {2m}{\hbar^2} [V_0(r) + 
V_{LS}(r)\langle {\bf l.s} \rangle_{lj}-\varepsilon(1+\varepsilon/2m)].
\label{eqa0}
\end{eqnarray}
The functions $P, Q, R, S$ can be expressed as  second order
 polynomials in $\kappa$
, namely,
$P=P_0+P_1k+P_2k^2$ and similarly for $Q, R, S$. Here 
$\kappa = (2j + 1)(l - j) = 
\frac{1}{2}\Omega (2j + 1)$ 
where $\Omega= +1$  if $j = l-\frac{1}{2}$
and $\Omega= -1$ if $j = l+\frac{1}{2}$.
It is easy to relate powers of 
$\kappa$ to $\langle {\bf l.s} \rangle_{lj}$. For instance, we have:
\begin{eqnarray}
2 \langle {\bf l.s} \rangle_{lj} & = & -{\hbar^2}(1 + \kappa~). 
\label{eqa00}
\end{eqnarray}
Thus, all contributions to $V_{LS}(r)$ can be identified by expressing the 
$\kappa$-dependence into a $\langle {\bf l.s} \rangle_{lj}$-dependence. One 
then obtains:
\begin{eqnarray}
V_{LS}(r) & = & V_{LS}^{(I)} + V_{LS}^{(II)} + V_{LS}^{(III)}~, 
\label{eqb1}
\end{eqnarray}
where
\begin{eqnarray}
V_{LS}^{(I)}& = & \frac {1} {m} \Big\{-\frac {\Lambda{'}}{\Lambda}
\Big(\frac {1}{r}+P_1\Big)+\frac {1}{r^2}-P{'}_1+\frac {R_1}{\Lambda}
+\Big(\frac {1}{r}+P_1\Big)(\Sigma^D_T+S_0)
+\Big(\frac {1}{r}+S_1\Big)(\Sigma^D_T+P_0)
\nonumber\\[6mm]
& &+\frac {\Lambda{'}}{\Lambda} P_2+P{'}_2-\frac {R_2}{\Lambda}
-(\Sigma^D_T+P_0)S_2-\Big(\frac {1}{r}+P_1\Big)\Big(\frac {1}{r}+S_1\Big)
-P_2(\Sigma^D_T+S_0)
\nonumber\\[6mm]
& &+[l(l+1)+1]\Big[\Big(\frac {1}{r}+P_1\Big)S_2+\Big(\frac {1}{r}
+S_1\Big)P_2\Big]-[2l(l+1)+1]P_2S_2\Big\}~,     
\label{eqa1}
\end{eqnarray}
\begin{eqnarray}
V_{LS}^{(II)} & = & \frac {1} {m} \Big\{-\frac {\Lambda{'}}{\Lambda}
(P_1-S_1)-(P_0-S_0)(P_1-S_1)
\nonumber\\[6mm]
& & +\frac {\Lambda{'}}{\Lambda}(P_2-S_2)+\frac {1}{2}(P_1-S_1)^2
+(P_0-S_0)(P_2-S_2)
\nonumber\\[6mm]
& & -[l(l+1)+1](P_1-S_1)(P_2-S_2)
+[2l(l+1)+1]\frac {1}{2} (P_2-S_2)^2\Big\}~,     
\label{eqa2}
\end{eqnarray}
\begin{eqnarray}
V_{LS}^{(III)} & = & \frac {1} {m} \Big\{-\frac {\Lambda{'}}{2\Lambda}
(P_1-S_1)+\frac {1}{2}(P{'}_1-S{'}_1)-\frac {1}{2}(P_0-S_0)(P_1-S_1)
\nonumber\\[6mm]
& & +\frac {\Lambda{'}}{2\Lambda}
(P_2-S_2)-\frac {1}{2}(P{'}_2-S{'}_2)+\frac {1}{4}(P_1-S_1)^2
+\frac {1}{2}(P_0-S_0)(P_2-S_2)
\nonumber\\[6mm]
& & -[l(l+1)+1]\frac {1}{2}(P_1-S_1)(P_2-S_2)
+[2l(l+1)+1]\frac {1}{4} (P_2-S_2)^2\Big\}~.     
\label{eqa3}
\end{eqnarray}

\end{appendix}


\begin{table}
\caption{The quantities $\Delta T$, $\Delta V_0$, $\Delta V_{LS}$ and 
$\varepsilon_{1d3/2}-\varepsilon_{1d5/2}$  
calculated with RHF for protons in $^{40}$Ca and $^{48}$Ca (in MeV). }
\begin{tabular}{ccccc}
Nucleus & $\Delta T$ & $\Delta V_0$ & $\Delta V_{LS}$ & $\varepsilon_{1d3/2}-
\varepsilon_{1d5/2}$  \\
\tableline
$^{40}$Ca & -2.94 & 3.31 & 7.54 & 8.01 \\
$^{48}$Ca & -1.73 & 4.25 & 1.38 & 4.06 \\
\end{tabular}
\label{table2}
\end{table}                                               

\begin{figure}
\caption{ The $\Delta(1d_{3/2}-1d_{5/2})$ proton splittings in Ca  
isotopes calculated with SIII (dot-dashed line), SkI4 (long-dashed line), 
RMFT (dotted line) and RHF (full line) models. Experimental values in A=40 
and A=48 are shown: A [10], B [11], C [8] and D [9].
} 
\label{fig1}
\end{figure}

\begin{figure}
\caption{ The $\Delta(1p_{1/2}-1p_{3/2})$ proton splittings in O 
isotopes calculated with SIII (dot-dashed line), SkI4 (long-dashed line), 
RMFT (dotted line) and RHF (full line) models. The experimental value for 
A=16 is indicated. }
\label{fig2}
\end{figure}

\begin{figure}
\caption{ The $\Delta(2p_{1/2}-2p_{3/2})$ proton splittings in Sn 
isotopes calculated with SIII (dot-dashed line), SkI4 (long-dashed line), 
RMFT (dotted line) and RHF (full line) models. Experimental values for 
A $\ge$112 are shown.}
\label{fig3}
\end{figure}

\begin{figure}
\caption{ The $\Delta(1f_{5/2}-1f_{7/2})$ proton splittings in Sn 
isotopes calculated with SIII (dot-dashed line), SkI4 (long-dashed line), 
RMFT (dotted line) and RHF (full line) models.}
\label{fig4}
\end{figure}

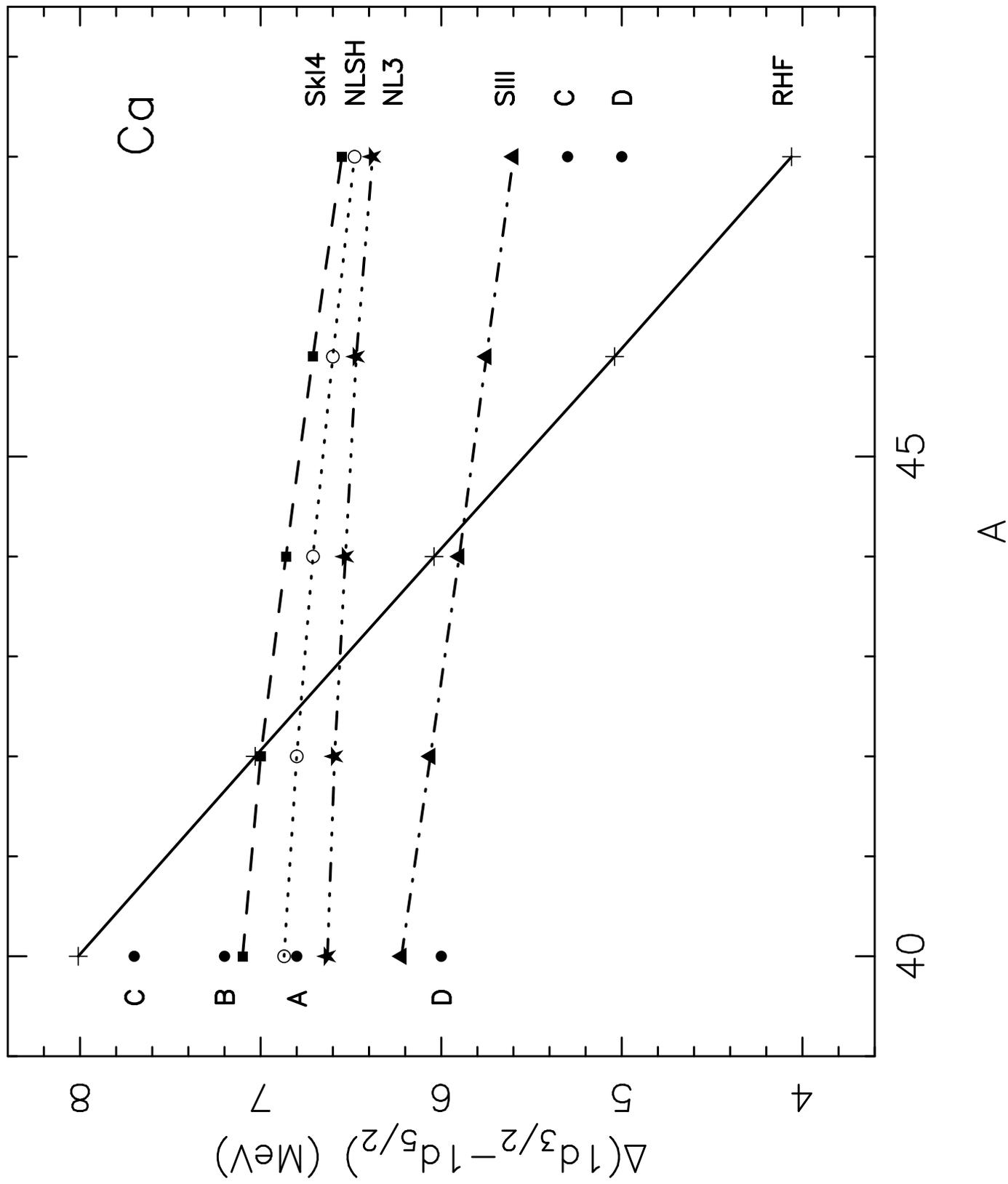
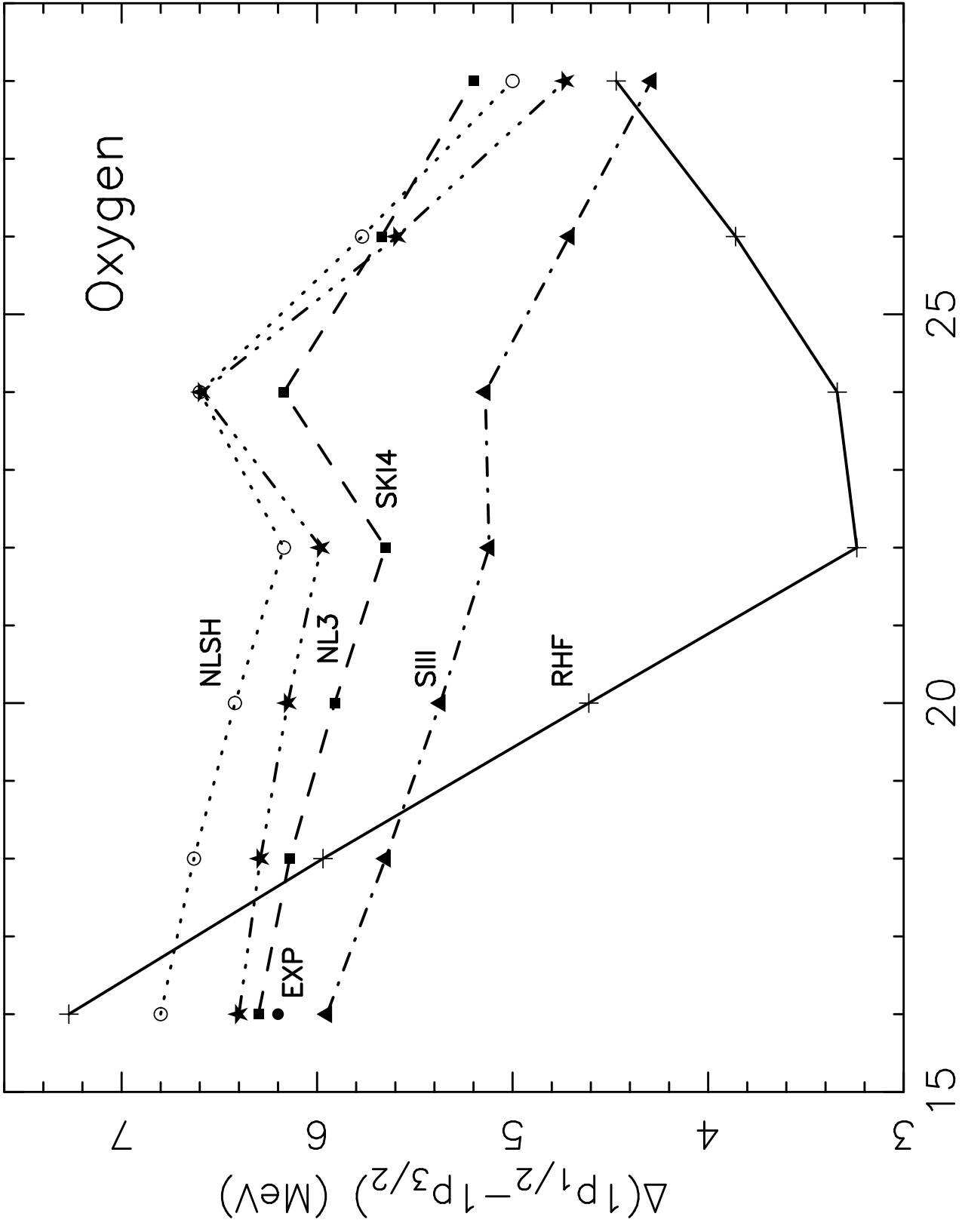
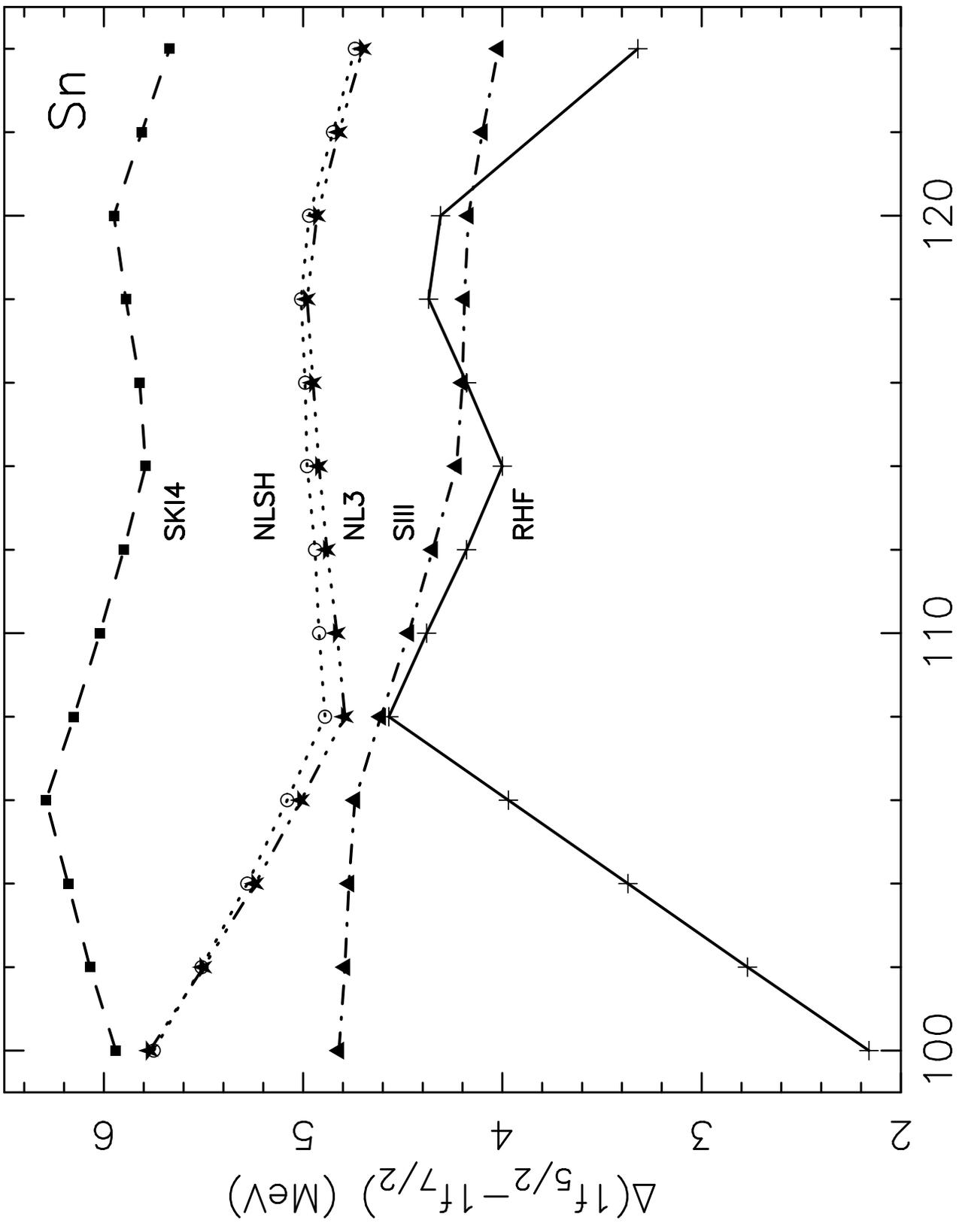
\end{document}